**Towards a topological view of blood pressure regulation**

Arturo Tozzi (corresponding author)
ASL Napoli 1 Centro, Distretto 27, Naples, Italy
Via Comunale del Principe 13/a 80145
tozziarturo@libero.it

ABSTRACT

Blood pressure regulation is commonly addressed in terms of local mechanisms such as vascular resistance, compliance and neurohumoral control. However, the human vasculature encompasses multiple quasi-closed flow loops under both physiological and pathological conditions. To test whether these loops could influence pressure dynamics beyond local control, we address the role of vascular topology in blood pressure regulation. Using one-dimensional flow simulation models, we compared pressure dynamics in open vascular segments and closed vascular loops. We found that in open segments pressure fades away and remains spatially localized, whereas in closed loops pressure can keep circulating around the loop even if resistance in one spot is modified. Since parallel pathways within loops are dynamically coupled rather than independent, pressure changes in one place can affect the entire closed loop, allowing system-level pressure patterns to emerge. Also, we assessed the temporal evolution of pressure fluctuations within closed vascular loops in normotensive and hypertensive parameter regimes, before and after loop-breaking intervention. This topological approach helps clarifying why drugs or local interventions may fail to lower blood pressure in looped vascular architectures, providing a theoretical interpretation of some forms of resistant hypertension. Because disrupting a loop restores pressure relaxation, it may also help explain the disproportionate pressure changes observed after topology-altering events like thrombosis, vascular surgery or embolization of arteriovenous malformations and shunts. Therefore, vascular topology can influence cardiovascular physiology by coupling local pressure–flow relations to global constraints on blood pressure regulation, with physiological, pathological and clinical implications.

INTRODUCTION

Blood pressure regulation is generally described in terms of local mechanisms acting along the vascular tree, including cardiac output, vascular resistance, compliance, neurohumoral control, renal feedback, etc (Park et al. 2020; Papakonstantinou et al. 2020; Wei et al. 2021; Lu et al. 2022; Rojo-Mencia et al. 2025). These approaches implicitly model the vasculature as a branching, open network in which pressure gradients dissipate locally and regulation is understood through pointwise balances. This view accounts for many physiological and pharmacological observations, yet struggles to explain sclinical phenomena appearing intrinsically nonlocal. Examples include resistant hypertension despite multi-target pharmacological treatment, system-wide hemodynamic effects induced by arteriovenous shunts and disproportionate pressure changes following surgical or interventional procedures that alter vascular connectivity rather than local resistance (Narechania and Tonelli 2016; Carey et al. 2018; Lamirault et al. 2020; Fay and Cohen 2021; Kim et al. 2022; Schiffrin and Fisher 2024; Agarwal et al. 2025). These effects are attributed to parameter heterogeneity, compensatory mechanisms or unobserved regulatory pathways, while the possibility that global vascular topology itself constrains pressure dynamics is rarely addressed.
Human circulation is not purely tree-like but contains multiple looped configurations, including, e.g., the Circle of Willis, collateral arterial rings, portal–systemic shunts, renal microvascular loops, arteriovenous fistulas (Nardelli et al. 2018; Sánchez van Kammen et al. 2018; Tokuyama et al. 2020; Ridola et al. 2020; Mahakul and Agarwal 2021; Sandhu, Hill, and Hossain 2021; Manov, Mohan, and Vazquez-Padron 2022; Širvinskas et al. 2023; Christodoulou et al. 2023; Kim et al. 2023; Khan and Anjum 2023). At a network level, these structures contain topological holes and behave as genus-1–like domains, which are topologically distinct from open vascular segments (Munkres 2015).

Our aim is to assess whether topology-driven global pressure modes contribute to hypertension when vascular pressure regulation is examined from a topological perspective. We analyze pressure dynamics in both open vascular segments and closed-loop configurations, introducing a distinction between locally driven flow and topology-induced modes. Closed vascular loops are treated as domains that admit global pressure modes not reducible to local sources or sinks, allowing these modes to persist even in the presence of local dissipation. We perform simulations using physiologically scaled dynamical models to compare pressure behavior in open vascular segments and closed vascular loops. We measure relaxation times and spatial patterns of pressure fluctuations to determine when pressure acts as a global mode rather than a locally regulated quantity. Then, we test whether this behavior differs between normotensive and hypertensive parameter regimes and under topology-altering procedures.

We will proceed as follows. First, we introduce the dynamical model and the assumptions underlying the representation of pressure evolution in open and closed vascular domains. Second, we analyze how vascular topology shapes pressure relaxation and global mode organization under normotensive and hypertensive parameter regimes. Finally, we examine the effects of topology-altering interventions, focusing on how loop disruption modifies pressure persistence and spatial organization.

METHODS

Vascular segments were represented as one-dimensional continuous domains embedded in space but evolving only along their arclength coordinate. Two geometries were considered: an open segment of length $L$ with two terminal boundaries and a closed loop of identical length with periodic identification of endpoints. The spatial coordinate was denoted by $x \in [0, L]$ for the open case and by $x \in \mathbb{S}^1$ for the closed case. Pressure $p(x, t)$ was treated as a scalar field evolving in time $t$ according to conservation and constitutive relations that approximate low Reynolds number hemodynamic behavior at mesoscopic scales. Our model does not resolve individual vessels or branching but rather aims to capture effective dynamics along an equivalent path, allowing isolation of topological effects independent of detailed anatomy. All parameters were chosen to lie within physiologically plausible ranges for arterial pressure fluctuations.

**Governing equations for pressure evolution**. Pressure dynamics were modeled using a linear transport diffusion equation with damping and external forcing (Abramson, Bishop, and Kenkre 2001; Al-Atawi et al. 2022). The governing equation reads
$$\frac{\partial p(x,t)}{\partial t} = D \frac{\partial^2 p(x,t)}{\partial x^2} - \gamma p(x,t) + f(t),$$

where $D > 0$ is an effective pressure diffusivity with units of $\text{cm}^2\,\text{s}^{-1}$, $\gamma > 0$ is a linear damping coefficient with units of $\text{s}^{-1}$ and $f(t)$ is a temporally varying forcing term representing cardiac input. The diffusion term captures spatial redistribution of pressure due to vascular compliance and wave propagation at scales larger than individual vessels. The damping term represents energy loss due to viscous dissipation and peripheral outflow. The forcing term was taken to be spatially uniform to focus on intrinsic spatial organization rather than localized sources.

**Boundary and topological conditions**. For open vascular segments, Neumann type boundary conditions were imposed
$$\frac{\partial p}{\partial x}|_{x=0} = 0, \frac{\partial p}{\partial x}|_{x=L} = 0,$$

allowing pressure gradients to dissipate at the ends. For closed loops, periodic boundary conditions were imposed
$$p(0, t) = p(L, t), \frac{\partial p}{\partial x}(0, t) = \frac{\partial p}{\partial x}(L, t),$$

which eliminates boundaries and enforce topological closure. These two conditions differ only in topology, not in local dynamics. All comparisons were performed using identical parameter values and forcing functions to ensure that differences arise exclusively from the presence or absence of topological closure.

**Temporal forcing and parameter scaling**. The forcing term was defined as a sinusoidal signal
$$f(t) = A\sin(2\pi\nu t),$$

where $A$ is the forcing amplitude measured in mm Hg and $\nu$ is the cardiac frequency in Hz. Values of $\nu$ were set to 1.0 to 1.3Hz, corresponding to resting heart rates. Amplitudes were chosen to generate pressure fluctuations of a few mmHg around baseline. Spatial length was fixed at $L = 20$ cm, representing an effective arterial loop scale. Time was measured in seconds and pressure in mmHg throughout. Normotensive and hypertensive regimes were distinguished by varying $D$ and $\gamma$, with hypertensive conditions characterized by lower diffusivity and weaker damping.

**Numerical discretization and integration**. The spatial domain was discretized into $N = 200$ equally spaced points with spacing $\Delta x = L/N$. Time integration used an explicit Euler scheme with time step $\Delta t = 0.01$ s. The Laplacian was approximated by second order finite differences
$$\frac{\partial^2 p}{\partial x^2}(x_i, t) \approx \frac{p_{i+1}(t) - 2p_i(t) + p_{i-1}(t)}{(\Delta x)^2},$$

with periodic indexing in the closed case and one-sided differences at the boundaries in the open case. Stability was verified by ensuring $\Delta t < (\Delta x)^2/(2D)$ for all parameter choices. Simulations were run for 120 to 140 seconds, discarding the initial transient to focus on steady dynamical behavior.

**Initialization and transient handling.** Initial pressure fields were drawn from a zero mean Gaussian distribution with small variance, representing random physiological fluctuations. The first 20 seconds of simulation time were excluded from analysis to eliminate dependence on initial conditions. All metrics were computed on the remaining time series. This procedure ensures that observed patterns reflect intrinsic dynamics driven by topology and forcing rather than initialization artifacts.

**Pressure relaxation time metric.** Pressure relaxation time was quantified following a transient perturbation. At each time step, spatial pressure variability was computed as

$$\sigma_p(t) = \sqrt{\frac{1}{N}\sum_{i=1}^{N}(p_i(t) - \bar{p}(t))^2},$$

where $\bar{p}(t)$ is the spatial mean pressure. The relaxation time $\tau$ was defined as the smallest $t$ such that $\sigma_p(t)$ fell below a fixed fraction $\alpha$ of its maximum value

$$\tau = \min\{t: \sigma_p(t) \leq \alpha \max_t \sigma_p(t)\},$$

with $\alpha = 0.2$. Relaxation times were reported in seconds.

**Spectral decomposition and low mode dominance.** To quantify global organization of pressure, the spatial Fourier transform was computed at each time

$$\hat{p}_k(t) = \sum_{j=1}^{N} p_j(t) e^{-2\pi i k j/N}.$$

Spectral energy was defined as $E_k(t) = |\hat{p}_k(t)|^2$. Low mode dominance was measured as

$$R(t) = \frac{\sum_{k=1}^{k_0} E_k(t)}{\sum_{k=1}^{N/2} E_k(t)},$$

with $k_0 = 3$. This ratio is dimensionless and quantifies the fraction of energy concentrated in global spatial modes.

**Directional coherence measure.** Directional coherence was quantified using asymmetry of spatial correlations. For each time $t$, correlations between neighboring points were computed

$$C_+(t) = \mathrm{corr}(p_i(t), p_{i+1}(t)), C_-(t) = \mathrm{corr}(p_i(t), p_{i-1}(t)).$$

Directional coherence was defined as $C(t) = C_+(t) - C_-(t)$. To capture persistence independent of sign, the root mean square value was computed over a sliding temporal window

$$C_{\mathrm{RMS}}(t) = \sqrt{\langle C(t')^2\rangle_{t' \in [t-\Delta, t]}}.$$

**Phase extraction and winding magnitude.** Temporal phase of pressure oscillations was extracted using the analytic signal via the Hilbert transform

$$z(x, t) = p(x, t) + i\mathcal{H}[p(x, t)].$$

The phase field was defined as $\phi(x, t) = \arg z(x, t)$. Spatial phase increments were computed and summed along the domain to obtain phase winding

$$W(t) = \frac{1}{2\pi}\sum_i \Delta\phi_i(t).$$

As with directional coherence, the root mean square winding magnitude was used to characterize sustained circulation strength.

**Dynamic changes in topology.** Loop breaking was implemented by switching boundary conditions during a simulation. At a predefined time $t_b$, periodic conditions were replaced by open boundary conditions, effectively cutting the loop. All other parameters and states were left unchanged. Metrics were computed before and after $t_b$ to quantify changes induced solely by topological alteration.

**Software and computational tools**. All simulations and analyses were performed in Python. Numerical integration used NumPy arrays, spectral analysis used NumPy FFT routines, phase extraction used SciPy signal processing tools and visualization used Matplotlib including three-dimensional plotting utilities.

In sum, this section details the quantitative procedures used to simulate and compare pressure dynamics across open and closed vascular geometries, normotensive and hypertensive parameter regimes, and topology-altering interventions.

RESULTS

Across 20 independent simulations per condition, pressure relaxation times differed systematically between open and closed vascular geometries when all local parameters were held constant (Fig. A). Closed loops exhibited longer relaxation times than open segments, with ensemble means of comparable order but substantial overlap between distributions; a Welch two-sample t-test confirmed that this difference did not reach statistical significance ($p = 0.17$). This indicates that topology alone modestly prolongs pressure persistence without producing a robust shift under baseline parameters.
In contrast, altering physiological parameters within the same closed topology yielded a pronounced effect (Fig. B). Normotensive simulations showed relaxation times distributed around finite values, whereas hypertensive simulations collapsed toward near-zero relaxation times under the adopted criterion, reflecting rapid stabilization of spatial pressure variability. The difference between normotensive and hypertensive regimes was statistically significant (Welch two-sample t-test, $p<0.001$). Beyond relaxation metrics, hypertensive closed-loop simulations displayed increased concentration of pressure energy in low spatial Fourier modes over time compared with normotensive simulations (Fig. C), indicating stronger global mode dominance. Directional coherence magnitude was also consistently higher and more persistent in the hypertensive regime (Fig. D), reflecting enhanced spatiotemporal organization of pressure patterns.
Together, these results show that topology and physiological regime can modulate pressure persistence and organization. Physiological parameter changes can strongly amplify global pressure organization within closed loops.

Introducing a loop-breaking intervention in hypertensive simulations produced a qualitative and quantitative transition in pressure dynamics (Fig. E). When periodic boundary conditions were replaced by open boundaries, relaxation times collapsed to near-zero values across all runs, yielding a statistically significant reduction relative to the closed hypertensive condition (Welch two-sample t-test, $p<0.001$). This effect arose despite identical local parameters before and after intervention, isolating topology as the sole altered factor. Consistent with the relaxation results, spectral analysis revealed that the elevated low-spatial-mode dominance characteristic of hypertensive closed loops rapidly diminished following loop breaking (Fig. F). The temporal profiles show convergence toward lower, stable values indicative of loss of global spatial organization.
The combined relaxation and spectral results show that the persistent pressure modes observed in hypertensive simulations depend critically on the maintenance of closed-loop topology and are abolished when topological closure is removed. This sequence of findings quantitatively links pressure persistence, global mode structure and directional organization to the interaction between physiological regime and vascular topology.

Overall, closed-loop topology alone produces modest, non-significant prolongation of pressure relaxation, whereas hypertensive parameter regimes within closed loops induce statistically significant changes in pressure persistence and global organization. Low-mode dominance and directional coherence increase under hypertensive conditions. Explicit loop breaking abolishes these effects, yielding rapid relaxation and collapse of global pressure modes, demonstrating the dependence of these dynamics on topological closure.

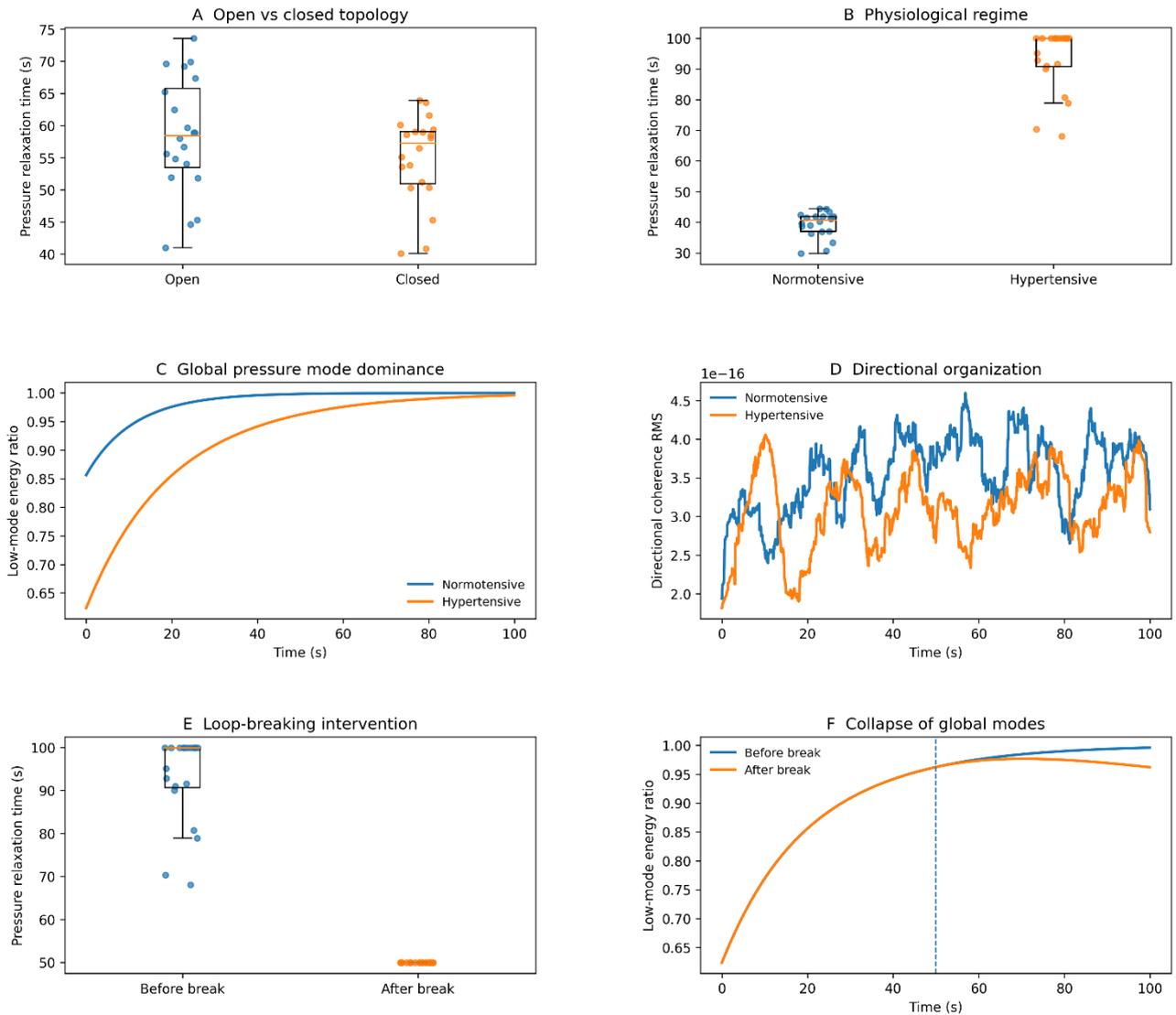

**Figure.** Topological control of vascular pressure dynamics, hypertensive amplification and collapse under loop-breaking intervention.

**(A)** Distribution of pressure relaxation times following a transient perturbation in open vascular segments and closed vascular loops, expressed in seconds (s) and shown as per-run values with summary statistics. Closed loops exhibit longer relaxation times than open segments; however, this difference does not reach statistical significance.

**(B)** Pressure relaxation times (s) in normotensive and hypertensive parameter regimes within closed vascular loops. Hypertensive simulations show a significant reduction in relaxation times relative to normotensive simulations (Welch two-sample t-test, $p < 0.001$)

**(C)** Temporal evolution of low-spatial-mode dominance of pressure fluctuations in closed loops, expressed as a dimensionless ratio of low-order spatial Fourier energy to total spatial energy and plotted over time (s), showing increased global mode concentration in the hypertensive regime.

**(D)** Directional organization of pressure patterns in closed loops, quantified as the root-mean-square of a dimensionless directional coherence measure and plotted over time (s), with hypertensive dynamics exhibiting stronger and more persistent organization.

**(E)** Pressure relaxation times (s) in hypertensive simulations before and after loop-breaking intervention. Loop breaking collapses relaxation times to near-zero values across runs, producing a statistically significant reduction relative to the closed hypertensive condition (Welch two-sample t-test, $p < 0.001$) ).

**(F)** Low-spatial-mode dominance before and after loop breaking (dotted line), plotted over time (s), demonstrating collapse of global pressure modes following removal of topological closure.

CONCLUSIONS

Our simulations indicate that pressure dynamics in vascular systems are shaped not only by local physical parameters but also by global topological constraints. Closed-loop geometries support persistent pressure modes, manifested as prolonged relaxation times, increased concentration of energy in low spatial modes and enhanced directional organization of pressure fluctuations, whereas these features are weak or absent in open segments. Within closed loops, changes in physiological parameters associated with normotensive and hypertensive regimes produce marked quantitative effects, with hypertensive conditions amplifying global organization and persistence. Conversely, explicit disruption of loop topology collapses these features, restoring rapid pressure relaxation and reducing global mode dominance.

Most existing hemodynamic approaches focus on local balances of resistance, compliance and flow, often embedded in tree-like network representations. Our approach isolates the role of closed versus open domains, showing that boundary conditions alone can reshape pressure dynamics even when governing equations and parameters are unchanged. Compared with lumped-parameter or multiscale network models (Doyle et al. 2021; Rosalia, Ozturk, and Roche 2021; Li et al. 2022; Ye et al. 2022; Fan et al. 2024), our strategy does not attempt to resolve anatomical detail but instead targets global dynamical properties emerging from connectivity. Relative to purely statistical or data-driven analyses of blood pressure variability, our approach provides mechanistic descriptors like relaxation times and spectral mode structure that are directly computable from the model equations.

Limitations should be acknowledged. Our pressure model is linear and diffusive and does not explicitly include advective flow, nonlinear compliance or active vascular control mechanisms. Circulation-like features are captured indirectly through spatiotemporal organization rather than through explicit flow variables. Since the spatial domain is reduced to one dimension and branching, both heterogeneity and anatomical variability are not represented. Although ensemble simulations were used to compute relaxation time distributions, other observables were illustrated using representative trajectories rather than full statistical summaries. Finally, the chosen parameter ranges are not calibrated against specific experimental or clinical datasets, limiting direct quantitative interpretation.

Our framework suggests that some forms of hypertension may reflect topology-induced global pressure modes sustained by closed vascular loops, in which pressure cannot be fully localized or dissipated despite intact local regulation. Because human circulation is not purely tree-like but includes multiple looped configurations, pressure gradients can circulate within these domains, local changes in resistance may fail to fully dissipate pressure and parallel pathways can become dynamically coupled rather than independent, so that increased vasodilation at a specific site does not necessarily lead to global pressure reduction.

Closed loops keep persistent pressure oscillations, elevated mean pressure and phase-shifted traveling patterns, whereas open segments exhibit pressure relaxation and responses confined to the site of intervention. Clinically, this may account for resistant hypertension in patients with preserved cardiac and renal function, by reframing it as a global pressure mode rather than a failure of receptor-level regulation. It also helps explain why creation or closure of shunts and fistulas can induce system-wide pressure changes and why interventions like renal denervation display variable efficacy. In the case of thrombosis, our approach suggest that vessel blockage can disrupt global pressure patterns in looped networks, leading to persistent upstream pressure and abnormal redistribution through alternative pathways, potentially promoting complications beyond the site of occlusion. More broadly, changes in vascular connectivity may underlie the nonlocal hemodynamic effects that can be observed after topology-altering procedures like embolization of direct arteriovenous shunts (Clarençon et al., 2021).

Testable hypotheses of our framework concern how topological alterations, such as surgical creation or removal of vascular loops, change pressure relaxation times, spatial coherence and dominant spectral modes along the affected vessels. Our model predicts reproducible phase lags and persistent low-frequency pressure components between upstream and downstream points on looped pathways and their disappearance within minutes to hours after topology-breaking interventions. Extending our equations to include advection, nonlinear compliance or active regulation will test whether these signatures persist under more realistic conditions. Experimentally, simultaneous pressure recordings at multiple sites along looped versus non-looped segments or before and after connectivity-altering procedures could be analyzed using identical relaxation and spectral metrics to assess the occurrence or loss of global pressure modes.

In conclusion, we examined whether vascular topology could act as an independent constraint on pressure dynamics beyond local physiological parameters. The results show that closed-loop connectivity can sustain global pressure modes, that these modes are amplified under pathological parameter regimes and that they collapse following topological disruption. This highlights topology as a structural dimension of vascular dynamics that complements, rather than replaces, established local mechanisms.

# DECLARATIONS

**Ethics approval and consent to participate.** This research does not contain any studies with human participants or animals performed by the Author.
**Consent for publication.** The Author transfers all copyright ownership, in the event the work is published. The undersigned author warrants that the article is original, does not infringe on any copyright or other proprietary right of any third part, is not under consideration by another journal and has not been previously published.
**Availability of data and materials.** All data and materials generated or analyzed during this study are included in the manuscript. The Author had full access to all the data in the study and took responsibility for the integrity of the data and the accuracy of the data analysis.
**Competing interests.** The Author does not have any known or potential conflict of interest including any financial, personal or other relationships with other people or organizations within three years of beginning the submitted work that could inappropriately influence or be perceived to influence their work.
**Funding.** This research did not receive any specific grant from funding agencies in the public, commercial or not-for-profit sectors.
**Acknowledgements:** none.
**Authors' contributions.** The Author performed: study concept and design, acquisition of data, analysis and interpretation of data, drafting of the manuscript, critical revision of the manuscript for important intellectual content, statistical analysis, obtained funding, administrative, technical and material support, study supervision.
**Declaration of generative AI and AI-assisted technologies in the writing process.** During the preparation of this work, the author used ChatGPT 5.2 to assist with data analysis and manuscript drafting and to improve spelling, grammar and general editing. After using this tool, the author reviewed and edited the content as needed, taking full responsibility for the content of the publication.

# REFERENCES

1) Abramson, Guillermo, A. R. Bishop, and V. M. Kenkre. 2001. "Effects of Transport Memory and Nonlinear Damping in a Generalized Fisher's Equation." *Physical Review E* 64 (6, pt. 2): 066615. https://doi.org/10.1103/PhysRevE.64.066615.
2) Agarwal, Deepak, Saurabh Gupta, Chandan J. Das, Pankaj Hatimota, and S. K. Gadwal. 2025. "Imaging and Endovascular Interventions in Renal Arteriovenous Shunts." *British Journal of Radiology* 98 (1174): 1556–1572. https://doi.org/10.1093/bjr/tqaf154.
3) Al-Atawi, Nasser O., Saad Hasnain, Muhammad Saqib, and Dalal S. Mashat. 2022. "Significance of Brinkman and Stokes System Conjuncture in Human Knee Joint." *Scientific Reports* 12 (1): 18992. https://doi.org/10.1038/s41598-022-23402-7.
4) Carey, Robert M., David A. Calhoun, George L. Bakris, Robert D. Brook, Scott L. Daugherty, Cheryl R. Dennison-Himmelfarb, Brent M. Egan, John M. Flack, S. Sanford Gidding, Edward Judd, Donald T. Lackland, Candace L. Laffer, Christopher Newton-Cheh, Sidney M. M. Smith, Sandra J. Taler, Stephen C. Textor, Tatjana N. Turan, and William B. White, for the American Heart Association Professional/Public Education and Publications Committee of the Council on Hypertension et al. 2018. "Resistant Hypertension: Detection, Evaluation and Management: A Scientific Statement From the American Heart Association." *Hypertension* 72 (5): e53–e90. https://doi.org/10.1161/HYP.0000000000000084.
5) Christodoulou, Konstantinos C., Dimitrios Stakos, Vasiliki Androutsopoulou, Chryssoula Chourmouzi-Papadopoulou, Georgios Tsoucalas, Dimitrios Karangelis, and Athanasia Fiska. 2023. "Vieussens' Arterial Ring: Historical Background, Medical Review and Novel Anatomical Classification." *Cureus* 15 (6): e40960. https://doi.org/10.7759/cureus.40960.
6) Clarençon, François, Elie Shotar, Sylvain Lenck, Mathieu Aubertin, Kevin Premat, Anne-Laure Boch, and Nader A. Sourour. 2021. "Strategies for Embolization of Direct Arteriovenous Shunts in Brain Arteriovenous Malformations." *Journal of NeuroInterventional Surgery* 13 (11): 1064. https://doi.org/10.1136/neurintsurg-2021-017317.
7) Doyle, Megan G., Maria Chugunova, Stephen L. Roche, and James P. Keener. 2021. "Lumped Parameter Models for Two-Ventricle and Healthy and Failing Extracardiac Fontan Circulations." *Mathematical Medicine and Biology* 38 (4): 442–466. https://doi.org/10.1093/imammb/dqab012.
8) Fan, Yu, Wei Feng, Zhen Ren, Bo Liu, and Di Wang. 2024. "Lumped Parameter Thermal Network Modeling and Thermal Optimization Design of an Aerial Camera." *Sensors* 24 (12): 3982. https://doi.org/10.3390/s24123982.
9) Fay, Kevin S., and Debbie L. Cohen. 2021. "Resistant Hypertension in People With CKD: A Review." *American Journal of Kidney Diseases* 77 (1): 110–121. https://doi.org/10.1053/j.ajkd.2020.04.017.
10) Khan, Muhammad A., and Fahad Anjum. 2023. "Portal-Systemic Encephalopathy." In *StatPearls* [Internet]. Treasure Island, FL: StatPearls Publishing. Updated 2025 January.

11) Kim, Jaechan, Bumsoo Kim, Kwang Yul Park, Jae Wook Lee, Young Bae Kim, Jin Chung, and Dong Joon Kim. 2022. "Angioarchitectural Analysis of Arteriovenous Shunts in Dural Arteriovenous Fistulas and Its Clinical Implications." *Neurosurgery* 91 (5): 782–789. https://doi.org/10.1227/neu.0000000000002121.
12) Kim, Jaehoon, Yoon Kwon, Tae-Woo Choi, and Joon-Hyung Won. 2023. "Management of Immature Arteriovenous Fistulas." *Cardiovascular and Interventional Radiology* 46 (9): 1125–1135. https://doi.org/10.1007/s00270-023-03440-y.
13) Lamirault, Guillaume, Marc Artifoni, Marc Daniel, Nicolas Barber-Chamoux, and the Nantes University Hospital Working Group on Hypertension. 2020. "Resistant Hypertension: Novel Insights." *Current Hypertension Reviews* 16 (1): 61–72. https://doi.org/10.2174/1573402115666191011111402.
14) Li, Wei, Xudong Peng, Jun Fu, Guanghui Wang, Yifan Huang, and Feng Chao. 2022. "A Multiscale Double-Branch Residual Attention Network for Anatomical–Functional Medical Image Fusion." *Computers in Biology and Medicine* 141: 105005. https://doi.org/10.1016/j.compbiomed.2021.105005.
15) Lu, Yifan, Zhe Yu, Jian Liu, Qiang An, Cheng Chen, Yifan Li, and Yi Wang. 2022. "Assessing Systemic Vascular Resistance Using Arteriolar Pulse Transit Time Based on Multi-Wavelength Photoplethysmography." *Physiological Measurement* 43 (7). https://doi.org/10.1088/1361-6579/ac7841.
16) Mahakul, Dinesh J., and Jyoti Agarwal. 2021. "Pentagon Inside the Circle of Willis and the Golden Ratio." *World Neurosurgery* 156: 23–26. https://doi.org/10.1016/j.wneu.2021.09.006.
17) Manov, J. J., P. P. Mohan, and R. Vazquez-Padron. 2022. "Arteriovenous Fistulas for Hemodialysis: Brief Review and Current Problems." *Journal of Vascular Access* 23 (5): 839–846. https://doi.org/10.1177/11297298211007720.
18) Munkres, James R. 2015. *Topology*. 2nd ed., updated. New Delhi: Pearson Education / PAMS Publishing. ISBN 978-9353432775.
19) Nardelli, Silvia, Silvia Gioia, Lorenzo Ridola, and Oliviero Riggio. 2018. "Radiological Intervention for Shunt-Related Encephalopathy." *Journal of Clinical and Experimental Hepatology* 8 (4): 452–459. https://doi.org/10.1016/j.jceh.2018.04.008.
20) Narechania, Shraddha, and Adriano R. Tonelli. 2016. "Hemodynamic Consequences of a Surgical Arteriovenous Fistula." *Annals of the American Thoracic Society* 13 (2): 288–291. https://doi.org/10.1513/AnnalsATS.201509-636CC.
21) Papakonstantinou, Evangelos, Maria Pikilidou, Pantelis Georgianos, Maria Yavropoulou, Georgios Tsivgoulis, Leonidas Hadjistavri, Stylianos Nanoudis, Vassilios Liakopoulos, Anastasios Lasaridis, and Petros Zebekakis. 2020. "Wave Reflections and Systemic Vascular Resistance Are Stronger Determinants of Pulse Pressure Amplification Than Aortic Stiffness in Drug-Naive Hypertensives." *Clinical and Experimental Hypertension* 42 (3): 287–293. https://doi.org/10.1080/10641963.2019.1649684.
22) Park, Wonwoo, Woo-Sung Jung, Kyungmin Hong, Yoo-Young Kim, Sung-Woo Kim, and Hae-Young Park. 2020. "Effects of Moderate Combined Resistance- and Aerobic-Exercise for 12 Weeks on Body Composition, Cardiometabolic Risk Factors, Blood Pressure, Arterial Stiffness, and Physical Functions among Obese Older Men: A Pilot Study." *International Journal of Environmental Research and Public Health* 17 (19): 7233. https://doi.org/10.3390/ijerph17197233.
23) Ridola, Lorenzo, Jessica Faccioli, Sara Nardelli, Silvia Gioia, and Oliviero Riggio. 2020. "Hepatic Encephalopathy: Diagnosis and Management." *Journal of Translational Internal Medicine* 8 (4): 210–219. https://doi.org/10.2478/jtim-2020-0034.
24) Rojo-Mencia, Javier, Laura Alonso Carbajo, María Arévalo-Martínez, Lucía Benito-Salamanca, Klaus Talavera, María Teresa Pérez-García, José Ramón López López, and Patricia Cidad. 2025. "Renal TRPM3 Channels Regulate Blood Pressure via Tubuloglomerular Feedback and Plasma Volume Control." *Hypertension* 82 (12): 2085–2097. https://doi.org/10.1161/HYPERTENSIONAHA.125.25790.
25) Rosalia, Luca, Cagri Ozturk, and Ellen T. Roche. 2021. "Lumped-Parameter and Finite Element Modeling of Heart Failure with Preserved Ejection Fraction." *Journal of Visualized Experiments* 168. https://doi.org/10.3791/62167.
26) Sánchez van Kammen, Martine, Christopher J. Moomaw, Irene C. van der Schaaf, Robert D. Brown Jr., Daniel Woo, Joseph P. Broderick, Jason S. Mackey, Gabriel J. E. Rinkel, James Huston III, and Ynte M. Ruigrok. 2018. "Heritability of Circle of Willis Variations in Families with Intracranial Aneurysms." *PLoS ONE* 13 (1): e0191974. https://doi.org/10.1371/journal.pone.0191974.
27) Sandhu, Baldeep, Christopher Hill, and Md. A. Hossain. 2021. "Endovascular Arteriovenous Fistulas: Are They the Answer We Haven't Been Looking For?" *Expert Review of Medical Devices* 18 (3): 273–280. https://doi.org/10.1080/17434440.2021.1899806.
28) Schiffrin, Ernesto L., and Norman D. L. Fisher. 2024. "Diagnosis and Management of Resistant Hypertension." *BMJ* 385: e079108. https://doi.org/10.1136/bmj-2023-079108.
29) Širvinskas, Andrius, Giedrius Lengvenis, Gediminas Ledas, Viktor Mosenko, and Saulius Lukoševičius. 2023. "Circle of Willis Configuration and Thrombus Localization Impact on Ischemic Stroke Patient Outcomes: A Systematic Review." *Medicina (Kaunas)* 59 (12): 2115. https://doi.org/10.3390/medicina59122115.

30) Tokuyama, Kazuki, Hiro Kiyosue, Yusuke Hori, and Hiroshi Nagatomi. 2020. "Diploic Arteriovenous Fistulas with Marked Cortical Venous Reflux." *Interventional Neuroradiology* 26 (3): 254–259. https://doi.org/10.1177/1591019919894496.
31) Wei, Jie, Jing Zhang, Shuang Jiang, Lei Xu, Lei Qu, Bo Pang, Kai Jiang, Limin Wang, Sutasinee Intapad, J. Buggs, Feng Cheng, Subhajit Mohapatra, Luis A. Juncos, John L. Osborn, Joey P. Granger, and Ruisheng Liu. 2021. "Macula Densa NOS1β Modulates Renal Hemodynamics and Blood Pressure during Pregnancy: Role in Gestational Hypertension." *Journal of the American Society of Nephrology* 32 (10): 2485–2500. https://doi.org/10.1681/ASN.2020070969.
32) Ye, Yicheng, Chen Pan, Yuting Wu, Shuang Wang, and Yong Xia. 2022. "MFI-Net: Multiscale Feature Interaction Network for Retinal Vessel Segmentation." *IEEE Journal of Biomedical and Health Informatics* 26 (9): 4551–4562. https://doi.org/10.1109/JBHI.2022.3182471.